\font\tenfrakturb=eufb10
\font\tenfraktur=eufm10
\font\tenmsbm=msbm10
\font\sevenfrakturb=eufb7
\font\sevenfraktur=eufm7
\font\sevenmsbm=msbm7
\font\fivefrakturb=eufb5
\font\fivefraktur=eufm5
\font\fivemsbm=msbm5
\newfam\bgothicfam
\newfam\gothicfam
\newfam\msbmfam
\textfont\bgothicfam = \tenfrakturb \scriptfont\bgothicfam=\sevenfrakturb
\scriptscriptfont\bgothicfam=\fivefrakturb

\textfont\gothicfam = \tenfraktur \scriptfont\gothicfam=\sevenfraktur
\scriptscriptfont\gothicfam=\fivefraktur

\textfont\msbmfam = \tenmsbm \scriptfont\msbmfam=\sevenmsbm
\scriptscriptfont\msbmfam=\fivemsbm

\def\Bbb{\tenmsbm\fam\msbmfam}
\def\gothb{\tenfrakturb\fam\bgothicfam}

\catcode`@=11
\def\renewcounter#1{\@definecounter{#1}\@ifnextchar[{\@newctr{#1}}{}}

\documentstyle[twoside]{article}
\catcode`\@=11
\long\def\@makefntext#1{
\protect\noindent \hbox to 3.2pt {\hskip-.9pt
$^{{\eightrm\@thefnmark}}$\hfil}#1\hfill} 

\def\@makefnmark{\hbox to 0pt{$^{\@thefnmark}$\hss}} 

\def\ps@myheadings{\let\@mkboth\@gobbletwo
\def\@oddhead{\hbox{}
\rightmark\hfil\eightrm\thepage}
\def\@oddfoot{}\def\@evenhead{\eightrm\thepage\hfil
\leftmark\hbox{}}\def\@evenfoot{}
\def\sectionmark##1{}\def\subsectionmark##1{}}
\oddsidemargin=\evensidemargin
\newcounter{sectionc}\newcounter{subsectionc}\newcounter{subsubsectionc}
\renewcommand{\section}[1] {\vspace{12pt}\addtocounter{sectionc}{1}
\setcounter{subsectionc}{0}\setcounter{subsubsectionc}{0}\noindent
	{\tenbf\thesectionc. #1}\par\vspace{5pt}}
\renewcommand{\subsection}[1] {\vspace{12pt}\addtocounter{subsectionc}{1}
	\setcounter{subsubsectionc}{0}\noindent
	{\bf\thesectionc.\thesubsectionc. {\kern1pt \bfit #1}}\par\vspace{5pt}}
\renewcommand{\subsubsection}[1] {\vspace{12pt}\addtocounter{subsubsectionc}{1}
	\noindent{\tenrm\thesectionc.\thesubsectionc.\thesubsubsectionc.
	{\kern1pt \tenit #1}}\par\vspace{5pt}}
\newcommand{\nonumsection}[1] {\vspace{12pt}\noindent{\tenbf #1}
	\par\vspace{5pt}}
\newcounter{appendixc}
\newcounter{subappendixc}[appendixc]
\newcounter{subsubappendixc}[subappendixc]
\renewcommand{\thesubappendixc}{\Alph{appendixc}.\arabic{subappendixc}}
\renewcommand{\thesubsubappendixc}
	{\Alph{appendixc}.\arabic{subappendixc}.\arabic{subsubappendixc}}
\renewcommand{\appendix}[1] {\vspace{12pt}
        \refstepcounter{appendixc}
        \setcounter{figure}{0}
        \setcounter{table}{0}
        \setcounter{lemma}{0}
        \setcounter{theorem}{0}
        \setcounter{corollary}{0}
        \setcounter{definition}{0}
        \setcounter{equation}{0}
        \renewcommand{\thefigure}{\Alph{appendixc}.\arabic{figure}}
        \renewcommand{\thetable}{\Alph{appendixc}.\arabic{table}}
        \renewcommand{\theappendixc}{\Alph{appendixc}}
        \renewcommand{\thelemma}{\Alph{appendixc}.\arabic{lemma}}
        \renewcommand{\thetheorem}{\Alph{appendixc}.\arabic{theorem}}
        \renewcommand{\thedefinition}{\Alph{appendixc}.\arabic{definition}}
        \renewcommand{\thecorollary}{\Alph{appendixc}.\arabic{corollary}}
        \renewcommand{\theequation}{\Alph{appendixc}.\arabic{equation}}
        \noindent{\tenbf Appendix \theappendixc #1}\par\vspace{5pt}}
\newcommand{\subappendix}[1] {\vspace{12pt}
        \refstepcounter{subappendixc}
        \noindent{\bf Appendix \thesubappendixc. {\kern1pt \bfit #1}}
	\par\vspace{5pt}}
\newcommand{\subsubappendix}[1] {\vspace{12pt}
        \refstepcounter{subsubappendixc}
        \noindent{\rm Appendix \thesubsubappendixc. {\kern1pt \tenit #1}}
	\par\vspace{5pt}}
\newcommand{\textlineskip}{\baselineskip=13pt}
\newcommand{\smalllineskip}{\baselineskip=10pt}
\def\eightcirc{
\begin{picture}(0,0)
\put(4.4,1.8){\circle{6.5}}
\end{picture}}
\def\eightcopyright{\eightcirc\kern2.7pt\hbox{\eightrm c}}
\newcommand{\copyrightheading}[1]
	{\vspace*{-2.5cm}\smalllineskip{\flushleft
	{\footnotesize Modern Physics Letters A, #1}\\
	{\footnotesize $\eightcopyright$\, World Scientific Publishing
	 Company}\\
         }}
\newcommand{\pub}[1]{{\begin{center}\footnotesize\smalllineskip
	Received #1\\
	\end{center}
        }}

\def\abstracts#1#2#3{{
        \centering{\begin{minipage}{4.5in}\baselineskip=10pt\footnotesize
        \parindent=0pt #1\par
        \parindent=15pt #2\par
        \parindent=15pt #3
        \end{minipage}}\par}}

\newcommand{\bibit}{\nineit}
\newcommand{\bibbf}{\ninebf}
\renewenvironment{thebibliography}[1]
         {\frenchspacing
         \ninerm\baselineskip=11pt
         \begin{list}{\arabic{enumi}.}
         {\usecounter{enumi}\setlength{\parsep}{0pt}
         \setlength{\leftmargin 12.7pt}{\rightmargin 0pt} 
         \setlength{\itemsep}{0pt} \settowidth
         {\labelwidth}{#1.}\sloppy}}{\end{list}}
\newcounter{itemlistc}
\newcounter{romanlistc}
\newcounter{alphlistc}
\newcounter{arabiclistc}

\newcommand{\fcaption}[1]{
         \refstepcounter{figure}
         \setbox\@tempboxa = \hbox{\footnotesize Fig.~\thefigure. #1}
         \ifdim \wd\@tempboxa > 5in
           {\begin{center}
         \parbox{5in}{\footnotesize\smalllineskip Fig.~\thefigure. #1}
            \end{center}}
        \else
             {\begin{center}
             {\footnotesize Fig.~\thefigure. #1}
              \end{center}}
        \fi}
\newcommand{\tcaption}[1]{
        \refstepcounter{table}
        \setbox\@tempboxa = \hbox{\footnotesize Table~\thetable. #1}
        \ifdim \wd\@tempboxa > 5in
           {\begin{center}
        \parbox{5in}{\footnotesize\smalllineskip Table~\thetable. #1}
            \end{center}}
        \else
             {\begin{center}
             {\footnotesize Table~\thetable. #1}
              \end{center}}
        \fi}
\def\@citex[#1]#2{\if@filesw\immediate\write\@auxout
        {\string\citation{#2}}\fi
\def\@citea{}\@cite{\@for\@citeb:=#2\do
        {\@citea\def\@citea{,}\@ifundefined
        {b@\@citeb}{{\bf ?}\@warning
        {Citation `\@citeb' on page \thepage \space undefined}}
        {\csname b@\@citeb\endcsname}}}{#1}}
\newif\if@cghi
\def\cite{\@cghitrue\@ifnextchar [{\@tempswatrue
        \@citex}{\@tempswafalse\@citex[]}}
\def\citelow{\@cghifalse\@ifnextchar [{\@tempswatrue
        \@citex}{\@tempswafalse\@citex[]}}
\def\@cite#1#2{{$\null^{#1}$\if@tempswa\typeout
        {IJCGA warning: optional citation argument
        ignored: `#2'} \fi}}

\def\pmb#1{\setbox0=\hbox{#1}
        \kern-.025em\copy0\kern-\wd0
        \kern.05em\copy0\kern-\wd0
        \kern-.025em\raise.0433em\box0}

\def\fnt#1#2{\footnotetext{\kern-.3em
        {$^{\mbox{\scriptsize #1}}$}{#2}}}
\def\fpage#1{\begingroup
\voffset=.3in
\thispagestyle{empty}\begin{table}[b]\centerline{\footnotesize #1}
       \end{table}\endgroup}
\def\runninghead#1#2{\pagestyle{myheadings}
\markboth{{\protect\footnotesize\it{\quad #1}}\hfill}
{\hfill{\protect\footnotesize\it{#2\quad}}}}
\font\tenrm=cmr10
\font\tenit=cmti10
\font\tenbf=cmbx10
\font\bfit=cmbxti10 at 10pt
\font\ninerm=cmr9
\font\nineit=cmti9
\font\ninebf=cmbx9
\font\eightrm=cmr8

\def\qed{\hbox{${\vcenter{\vbox{  
   \hrule height 0.4pt\hbox{\vrule width 0.4pt height 6pt
   \kern5pt\vrule width 0.4pt}\hrule height 0.4pt}}}$}}

\begin{document}
\runninghead{Yu. P. Goncharov}
{Black hole physics, confining solutions of SU(3)-Yang-Mills equations and...}
\normalsize\textlineskip
\thispagestyle{empty}
\setcounter{page}{1}
\copyrightheading{Vol. 16, No. 9 (2001) 557-569}
\vspace*{0.88truein}
\fpage{1}
\centerline{\bf BLACK HOLE PHYSICS, CONFINING SOLUTIONS OF SU(3)-YANG-MILLS }
\vspace*{0.035truein}
\centerline{\bf EQUATIONS AND RELATIVISTIC MODELS OF MESONS}
\vspace*{0.035truein}
\vspace*{0.37truein}
\centerline{\footnotesize YU. P. GONCHAROV}
\vspace*{0.015truein}
\centerline{\footnotesize\it Theoretical Group,
Experimental Physics Department, State Technical University}
\baselineskip=10pt
\centerline{\footnotesize\it Sankt-Petersburg 195251, Russia}
\vspace*{0.225truein}
\pub{December 2000}
\vspace*{0.21truein}
\abstracts{
  The black hole physics techniques and results are applied to find the set
of the exact solutions of the SU(3)-Yang-Mills equations in
Minkowski spacetime in the Lorentz gauge. All the solutions contain only
the Coulomb-like or linear in $r$ components of SU(3)-connection. This
allows one to obtain some possible
exact and approximate solutions of the corresponding Dirac equation
that can describe the relativistic bound states. Possible application to
the relativistic models of mesons is also outlined.
}{}{}
\vspace*{1pt}\textlineskip 
\section{Introductory Remarks} 
\vspace*{-0.5pt}
\noindent

  The present paper is motivated by the part of hadronic physics studying
mesons. As is known, mesons within the hadron physics are the one of
central sources of information about the quark interaction. Since nearly
all known mesons are bound states of a quark $q$ and an antiquark
$\overline{q'}$ (where the flavours of $q$ and $\overline{q'}$ may be
different) then actually all modern meson spectroscopy is based on one or
another quark model of mesons.
Referring for more details, e. g., to the recent up-to-date
review\cite{Fel00} and references therein, it should
be noted here that at present some generally accepted relativistic model of
mesons is absent. The description of mesons in quark models is actually
implemented by nonrelativistic manner (for example, on the basis of
the Schr{\"o}dinger equation) and
then one tries to include relativistic corrections in one or another way.
Such an inclusion is not single-valued and varies in dependence of the point
of view for different authors (see, e. g. Refs.\cite{Br98} and references
therein). It would be more consistent, to our mind, building a primordially
relativistic model so that one can then pass on to the nonrelativistic one by
the standard limiting transition and, as a result, to estimate the relativistic
effects in self-consistent way.

 As follows from the main principles of quantum chromodynamics (QCD),
the appropriate relativistic models for description of mesons as relativistic
bound states of quarks should consist in considering the solutions of Dirac
equation in a SU(3)-Yang-Mills field representing gluonic field. The latter
should be the solution of the corresponding Yang-Mills equations and
should model the quark confinement. Such solutions are usually supposed
to contain at least
the components of the mentioned SU(3)-field which are Coulomb-like or linear
in $r$, the distance between
quarks, and in what follows we call these solutions the {\it confining} ones.
No general recipe of obtaining such solutions exists at present.
In the given paper for this aim we shall employ
the techniques used in Refs.\cite{Gon678} for finding the U(N)-monopole
solutions in black hole physics and, to analyse the corresponding Dirac
equation, we shall use the results of Refs.\cite{Gon99} about spinor fields
on black holes.

Further we shall deal with the metric of
the flat Minkowski spacetime $M$ that
we write down (using the ordinary set of local spherical coordinates
$r,\vartheta,\varphi$ for spatial part) in the form
$$ds^2=g_{\mu\nu}dx^\mu\otimes dx^\nu\equiv
dt^2-dr^2-r^2(d\vartheta^2+\sin^2\vartheta d\varphi^2)\>. \eqno(1)$$
As is known, such a metric can be obtained from the Schwarzschild black hole
metric when the black hole mass is equal to 0.
Besides we have
$|\delta|=|\det(g_{\mu\nu})|=(r^2\sin\vartheta)^2$
and $0\leq r<\infty$, $0\leq\vartheta<\pi$,
$0\leq\varphi<2\pi$.

  Throughout the paper we employ the system of units with $\hbar=c=G=1$.
Finally, we shall denote $L_2(F)$ the set of the modulo square integrable
complex functions on any manifold $F$ furnished with an integration measure
while $L^n_2(F)$ will be the $n$-fold direct product of $L_2(F)$
endowed with the obvious scalar product.

\section{Dirac equation}
  To formulate the Dirac equation needed to us, let us notice that
the relativistic wave function of meson can be chosen in the form
$$\psi=\pmatrix{\psi_1\cr\psi_2\cr\psi_3\cr}$$
with the four-dimensional spinors $\psi_j$ representing $j$-th colour
component of meson.
Under this situation, if denoting
$S(M)$ and $\xi$, respectively, the standard spinor bundle
and three-dimensional vector one (equipped with a SU(3)-connection) over
Minkowski spacetime, we can
construct tensorial product $S(M)\otimes\xi$. It is clear that $\psi$ is just
a section of the latter bundle. Then the corresponding Dirac equation
for $\psi$ can be obtained with using the construction of Dirac operator
${\cal D}$ with coefficients in $\xi$ and may look as follows

$${\cal D}\psi=\mu_0\psi,\>\eqno(2)$$
where $\mu_0$ is the reduced relativistic mass which is equal, e. g., for
quarkonia, to one half a constituent mass of quarks forming quarkonium,
but definition of $\mu_0$ is, generally
speaking, not single-valued and requires the specification
within the concrete problem (see Ref.\cite{DTF}). Under this sitiation the
coordinate $r$ makes sense of the distance between quarks.

 From general considerations\cite{{81},{89},{Bes87}} the explicit form of
the operator ${\cal D}$ in local coordinates $x^\mu$ on a $2k$-dimensional
(pseudo)riemannian manifold can be written as follows
$${\cal D}=i(\gamma^e\otimes I_3)E_e^\mu(\partial_\mu\otimes I_3
-\frac{1}{2}\omega_{\mu ab}\gamma^a\gamma^b\otimes I_3-igA_\mu),
\>a < b ,\>\eqno(3)$$
where $A=A_\mu dx^\mu$, $A_\mu=A^c_\mu T_c$ is a SU(3)-connection in the
bundle $\xi$, $I_3$ is the unit matrix $3\times3$, the matrices $T_c$ form
a basis of the Lie algebra of SU(3) in 3-dimensional space (we consider
$T_a$ hermitean which is acceptable in physics), $c=1,...,8$, $\otimes$ here
means tensorial product of matrices, $g$ is a gauge coupling constant.
Further, the forms
$\omega_{ab}=\omega_{\mu ab}dx^\mu$ obey the Cartan structure equations
$de^a=\omega^a_{\ b}\wedge e^b$ with exterior derivative $d$, while the
orthonormal basis $e^a=e^a_\mu dx^\mu$ in cotangent bundle and
dual basis $E_a=E^\mu_a\partial_\mu$ in tangent bundle are connected by the
relations $e^a(E_b)=\delta^a_b$. At last, matrices $\gamma^a$ represent
the Clifford algebra of
the corresponding quadratic form in ${\Bbb C}^{2^k}$. Below we shall deal only
with 4D lorentzian case (quadratic form $Q_{1,3}=x_0^2-x_1^2-x_2^2-x_3^2$).
For this we take the following choice for $\gamma^a$
$$\gamma^0=\pmatrix{1&0\cr 0&-1\cr}\,,
\gamma^b=\pmatrix{0&\sigma_b\cr-\sigma_b&0\cr}\,,
b= 1,2,3\>, \eqno(4)$$
where $\sigma_b$ denote the ordinary Pauli matrices.
It should be noted that, in lorentzian case, Greek indices $\mu,\nu,...$
are raised and lowered with $g_{\mu\nu}$ of (1) or its inverse $g^{\mu\nu}$
and Latin indices $a,b,...$ are raised and lowered by
$\eta_{ab}=\eta^{ab}$= diag(1,-1,-1,-1),
so that $e^a_\mu e^b_\nu g^{\mu\nu}=\eta^{ab}$,
$E^\mu_aE^\nu_bg_{\mu\nu}=\eta_{ab}$ and so on.

Using the fact that all bundles over Minkowski spacetime are trivial
and, as a result, they can be
trivialized over the chart of local coordinates
$(t,r,\vartheta,\varphi)$ covering almost the whole Minkowski manifold,
we can concretize the Dirac equation (2) on the given chart for
$\psi$ in the case of metric (1).
Namely, we can put $e^0=dt$, $e^1=dr$,
$e^2=rd\vartheta$, $e^3=r\sin{\vartheta}d\varphi$ and, accordingly,
$E_0=\partial_t$, $E_1=\partial_r$,
$E_2=\partial_\vartheta/r$, $E_3=\partial_\varphi/(r\sin{\vartheta})$.
This entails
$$\omega_{12}=-d\vartheta,
\omega_{13}=-\sin{\vartheta}d\varphi,
\omega_{23}=-\cos{\vartheta}d\varphi.\>\eqno(5)$$
As for the connection $A_\mu$ in bundle $\xi$ then the suitable one should be
the confining solution of the Yang-Mills equations
$$dF=F\wedge A - A\wedge F \>, \eqno(6)$$
$$d\ast F= \ast F\wedge A - A\wedge\ast F +J\>\eqno(7)$$
with the exterior differential $d=\partial_t dt+\partial_r dr+
\partial_\vartheta d\vartheta+\partial_\varphi d\varphi$ in coordinates
$t,r,\vartheta,\varphi$ while the curvature matrix (field strentgh)
for $\xi$-bundle is $F=dA+A\wedge A$ and $\ast$ means the Hodge star
operator conforming to metric (1), $J$ is a source.
It is clear that (6) is identically satisfied --- this
is just the Bianchi identity holding true for any connection (see,
e. g., Refs.\cite{Bes87}) so that it is necessary to solve only the 
equations (7).
Introducing the Hodge star operator $\ast$
on 2-forms $F= F^a_{\mu\nu}T_adx^\mu\wedge dx^\nu$ with the values
in the Lie algebra of ${\rm SU(3)}$ by the relation (see, e. g.,
Ref.\cite{81})
$$(F^a_{\mu\nu} dx^\mu\wedge dx^\nu)\wedge
(\ast F^a_{\alpha\beta} dx^\alpha\wedge
dx^\beta)=(g^{\mu\alpha}g^{\nu\beta}-g^{\mu\beta}g^{\nu\alpha})
F^a_{\mu\nu}F^a_{\alpha\beta}
\sqrt{\delta}\,dx^0\wedge\cdots\wedge dx^3 \>\eqno(8)$$
written in local coordinates $x^\mu$ [there is no summation over $a$
in (8)], in coordinates $t,r,\vartheta,\varphi$ we have the relations
$$\ast(dt\wedge dr)=-r^2\sin\vartheta d\vartheta\wedge d\varphi\>,
\ast(dt\wedge d\vartheta)=\sin\vartheta dr\wedge d\varphi\>,$$
$$\ast(dt\wedge d\varphi)=-\frac{1}{\sin\vartheta}dr\wedge d\vartheta\>,
\ast(dr\wedge d\vartheta)=\sin\vartheta dt\wedge d\varphi\>,$$
$$\ast(dr\wedge d\varphi)=-\frac{1}{\sin\vartheta}dt\wedge d\vartheta\>,
\ast(d\vartheta\wedge d\varphi)=\frac{1}{r^2\sin\vartheta}dt\wedge dr\>,
\eqno(9)$$
so that $\ast^2=\ast\ast=-1$, as should be for the manifolds with lorentzian
signature.\cite{Bes87}

Besides the sought solutions are usually believed to obey an additional
gauge condition.
In the present paper we take the Lorentz gauge condition that can be written
in the form ${\rm div}(A)=0$, where the divergence of the Lie algebra valued
1-form $A=A^c_\mu T_cdx^\mu$ is defined by the relation (see, e. g.
Refs.\cite{Bes87})
$${\rm div}(A)=\frac{1}{\sqrt{|\delta|}}\partial_\mu(\sqrt{|\delta|}g^{\mu\nu}
A_\nu)\>.\eqno(10)$$

\section{Confining solutions}
 Now we can use the techniques of Refs.\cite{Gon678} to find a set of the
confining solutions of Eq. (7). The essence of those techiniques consists in
systematic usage of the Hodge star operator. Let firstly $J=0$ and let us
put $T_c=\lambda_c$,
where $\lambda_c$ are the Gell-Mann matrices (whose explicit form can be
found in Refs.\cite{Gon678}), and in detail let us write out the addend
$A_\mu=A^c_\mu T_c$ of the operator ${\cal D}$ of (3)
$$A^c_\mu\lambda_c=
\pmatrix
{A^3_\mu+\frac{1}{\sqrt{3}}A^8_\mu&A^1_\mu-iA^2_\mu&A^4_\mu-iA^5_\mu\cr
A^1_\mu+iA^2_\mu&-A^3_\mu+\frac{1}{\sqrt{3}}A^8_\mu&A^6_\mu-iA^7_\mu\cr
A^4_\mu+iA^5_\mu&A^6_\mu+iA^7_\mu&-\frac{2}{\sqrt{3}}A^8_\mu\cr}\>.
\eqno(11)$$
Then it is naturally to put $A^c_\mu=0$ with $c=1,2,4,5,6,7$ and to obtain
from (2) the following system of Dirac equations for colour
components $\psi_j$
$$i\gamma^eE^\mu_e\left[\partial_\mu
-\frac{1}{2}\omega_{\mu ab}\gamma^a\gamma^b-ig\left(A^3_\mu+
\frac{1}{\sqrt{3}}A^8_\mu\right)\right]\psi_1 =\mu_0\psi_1 \>,\eqno(12)$$
$$i\gamma^eE^\mu_e\left[\partial_\mu
-\frac{1}{2}\omega_{\mu ab}\gamma^a\gamma^b-ig\left(-A^3_\mu+
\frac{1}{\sqrt{3}}A^8_\mu\right)\right]\psi_2 =\mu_0\psi_2 \>,\eqno(13)$$
$$i\gamma^eE^\mu_e\left[\partial_\mu
-\frac{1}{2}\omega_{\mu ab}\gamma^a\gamma^b-
ig\left(-\frac{2}{\sqrt{3}}A^8_\mu\right)\right]\psi_3=\mu_0\psi_3
\>.\eqno(14)$$
Further for to avoid unnecessary complications with the Lorentz condition,
it is simpler of all to put $A^{3,8}_{r,\vartheta}=0$. After this we search
for the solution of (7) in the form $A=A_t(r)dt+A_\varphi(r)d\varphi$
with $A_{t,\varphi}=A^3_{t,\varphi}\lambda_3+A^8_{t,\varphi}\lambda_8$.
It is then easy to check that $A\wedge A=0$, $F=dA=
-\partial_rA_tdt\wedge dr+\partial_rA_\varphi dr\wedge d\varphi$ and, with the
help of the relations (9),
$\ast F=r^2\sin\vartheta\partial_rA_td\vartheta\wedge d\varphi-
\frac{1}{\sin\vartheta}\partial_rA_\varphi dt\wedge d\vartheta$. From here
it follows that $\ast F\wedge A=A\wedge\ast F$ and Eq. (7) will be equivalent
to the equation $d\ast F=0$ while the latter yields
$$\partial_r(r^2\partial_rA_t)=0,\>\partial^2_rA_\varphi=0\>,\eqno(15)$$
and we write down the solutions of (15) in the combinations that are just
needed to insert into (12)--(14)
$$ A^3_t+\frac{1}{\sqrt{3}}A^8_t =-\frac{a_1}{r}+A_1 \>,$$
$$ -A^3_t+\frac{1}{\sqrt{3}}A^8_t=\frac{a_1+a_2}{r}-(A_1+A_2)\>,$$
$$-\frac{2}{\sqrt{3}}A^8_t=-\frac{a_2}{r}+A_2\>, \eqno(16)$$
$$ A^3_\varphi+\frac{1}{\sqrt{3}}A^8_\varphi =b_1r+B_1 \>,$$
$$ -A^3_\varphi+\frac{1}{\sqrt{3}}A^8_\varphi=-(b_1+b_2)r-(B_1+B_2)\>,$$
$$-\frac{2}{\sqrt{3}}A^8_\varphi=b_2r+B_2\> \eqno(17)$$
with some constants $a_j, A_j, b_j, B_j$ parametrizing solutions.

Another class of the confining solutions of (7) can be obtained with the aid
of the ansatz $A=A_t(r)dt+A_\varphi(r,\vartheta)d\varphi$, i. e. now we
consider the component $A_\varphi$ depending also on $\vartheta$.
Again we shall have $A\wedge A=0$, $F=dA=
-\partial_rA_tdt\wedge dr+\partial_rA_\varphi dr\wedge d\varphi
+\partial_\vartheta A_\varphi d\vartheta\wedge d\varphi$
and $\ast F=r^2\sin\vartheta\partial_rA_td\vartheta\wedge d\varphi-
\frac{1}{\sin\vartheta}\partial_rA_\varphi dt\wedge d\vartheta
+\frac{1}{r^2\sin\vartheta}\partial_\vartheta A_\varphi dt\wedge dr$. Then
anew $\ast F\wedge A=A\wedge\ast F$ and Eq. (7) is converted into
$d\ast F=0$ that entails
$$\partial_r(r^2\partial_rA_t)=0,\>\eqno(18)$$
$$r^2\partial^2_rA_\varphi
+\sin\vartheta\partial_\vartheta\left(\frac{1}{\sin\vartheta}
\partial_\vartheta A_\varphi\right)=0\>.\eqno(19)$$
It is clear that Eq. (18) gives the same solutions of (16). We shall not
here discuss the general form of the solution for Eq. (19) and only write
out the possible solution of it which is useful to us in the present
paper in the form slightly modifying (17)

$$ A^3_\varphi+\frac{1}{\sqrt{3}}A^8_\varphi =-K_1\cos\vartheta+b_1r+B_1 \>,$$
$$ -A^3_\varphi+\frac{1}{\sqrt{3}}A^8_\varphi=(K_1+K_2)\cos\vartheta-
(b_1+b_2)r-(B_1+B_2)\>,$$
$$-\frac{2}{\sqrt{3}}A^8_\varphi=-K_2\cos\vartheta+b_2r+B_2\> \eqno(20)$$
with some real constants $K_j, b_j, B_j$ parametrizing solution.

Finally, we shall adduce one solution of (7) with a source. It is given by
relations of (16) for $A_t$ while for $A_\varphi$ we have
$$A_\varphi=I_3[({\bf b}r+{\bf B})\sin\vartheta+{\bf K}\cos\vartheta]\>
\eqno(21)$$
with vectors ${\bf b}=(b_1,-b_1-b_2,b_2)$, ${\bf B}=(B_1,-B_1-B_2,B_2)$,
${\bf K}=(-K_1, K_1+K_2,-K_2)$. The conforming source is
$J=-\frac{1}{r^2\sin^2\vartheta}I_3({\bf b}r+{\bf B})
dt\wedge dr\wedge d\vartheta$ and since $*J={\gothb j}={\gothb j}_\mu dx^\mu=
{\gothb j}_\varphi d\varphi$ with ${\gothb j}_\varphi=
-\frac{1}{r^2\sin\vartheta}I_3({\bf b}r+{\bf B})$, ${\rm div}({\gothb j})=0$,
we can interpret ${\gothb j}$ as the colour current giving rise to the
corresponding linear interaction.

At last, it is easy to check that all the solutions obtained satisfy
the Lorentz gauge condition.

\section{Spectrum of bound states in the Coulomb-like case}
 Returning to the Eqs. (12)--(14), we should note that
when inserting the confining solutions gained into them the variables
$r$ and $\vartheta$ cannot be, generally speaking, separated. Let us take,
for example, the solutions of (16) and (21) for the Yang-Mills equations (7)
with the source $J$ described late in Sec. 3 (where we at first put $K_j=0$)
and for all three equations let us employ the ansatz
$$\psi_j=e^{i\omega_j t}r^{-1}\pmatrix{F_{j1}(r)\Phi_j(\vartheta,\varphi)\cr\
F_{j2}(r)\sigma_1\Phi_j(\vartheta,\varphi)}\>,j=1,2,3\eqno(22)$$
with a 2D spinor $\Phi_j=\pmatrix{\Phi_{j1}\cr\Phi_{j2}}$. Then, in the way
analogous to that of Refs.\cite{Gon99}, we can get, e. g., from (12) the
system
$$\left[\left(\partial_r+\frac{1}{r}\right)+\frac{1}{r}{\cal D}_0-\sigma_2
g\left(b_1+\frac{B_1}{r}\right)\right]\frac{1}{r}F_{11}\Phi_1=
i(\mu_0-c_1)\frac{1}{r}F_{12}\Phi_1,$$
$$\left[\left(\partial_r+\frac{1}{r}\right)+\frac{1}{r}{\cal D}_0-\sigma_2
g\left(b_1+\frac{B_1}{r}\right)\right]\frac{1}{r}F_{12}\sigma_1\Phi_1=
-i(\mu_0+c_1)\frac{1}{r}F_{11}\sigma_1\Phi_1\eqno(23)$$
with $c_1=\omega_1-g(-a_1/r+A_1)$ while
$${\cal D}_0=\sigma_1\sigma_2\partial_\vartheta+\frac{1}{\sin\vartheta}
\sigma_1\sigma_3\left(\partial_\varphi-
\frac{1}{2}\sigma_2\sigma_3\cos\vartheta\right)\eqno(24)$$
is the euclidean Dirac operator on the unit sphere ${\Bbb S}^2$. It is not
complicated to check that at $b_1\ne0$, $B_1\ne0$ the variables $r$ and
$\vartheta$ are not separated.
Under this situation
we shall at first restrict ourselves to the case
$b_j=B_j=0$ since under the circumstances we can use the results of
Refs.\cite{Gon99} for to solve Eqs. (12)--(14) exactly. Primarily let us
put also $K_j=0$ and then we shall modify results to take into account
the case $K_j\ne0$.
For all three equations we employ the ansatz (22) and in the way
analogous to that of Refs.\cite{Gon99}, we can obtain from (12)--(14)
the systems
$$\left(\partial_r+
\frac{\lambda_j}{r}\right)F_{j1}=
i(\mu_0-c_j)F_{j2},$$
$$\left(\partial_r
-\frac{\lambda_j}{r}\right)F_{j2}=
-i(\mu_0+c_j)F_{j1} \>\eqno(25)$$
with an eigenvalue $\lambda_j$ for the eigenspinor $\Phi_j$ of the above
operator ${\cal D}_0$,
$\lambda_j=\pm(l_j+1)\in{\Bbb Z}\backslash\{0\}\>, l_j=0,1,2...$
(for more details see Refs.\cite{Gon99}). Besides
$$c_2=\omega_2-g\left[\frac{a_1+a_2}{r}-(A_1+A_2)\right]\>,$$
$$c_3=\omega_3-g\left(-\frac{a_2}{r}+A_2\right)\>,\eqno(26)$$
so that the energy spectrum $\varepsilon$ of meson is given by the
relation $\varepsilon=\omega_1+\omega_2+\omega_3$.
The explicit form of $\Phi_j$ is not needed here and
can be found in
Refs.\cite{Gon99}. For the purpose of the present paper it is sufficient
to know that spinors $\Phi_j$ can be subject to
the normalization condition
$$\int\limits_0^\pi\,\int\limits_0^{2\pi}(|\Phi_{j1}|^2+|\Phi_{j2}|^2)
\sin\vartheta d\vartheta d\varphi=1\> , \eqno(27)$$
i. e., they form an orthonormal basis in $L_2^2({\Bbb S}^2)$. Further we
restrict ourselves to the case $j=1$ because for $j=2,3$ the considerations
are the same.

Let us employ the ansatz
$F_{11}=\sqrt{\mu_0-(\omega_1-gA_1)}r^{\alpha_1}e^{-\beta_1r}[f_{11}(r_1)+
f_{12}(r_1)]$,
$F_{12}=i\sqrt{\mu_0+(\omega_1-gA_1)}r^{\alpha_1}e^{-\beta_1r}[f_{11}(r_1)-
f_{12}(r_1)]$ with $\alpha_1=\sqrt{ \lambda_1^2-g^2a_1^2}$, $\beta_1=
\sqrt{\mu_0^2-(\omega_1-gA_1)^2}$, $r_1=2\beta_1r$.
Then, inserting the ansatz into (25), adding and subtracting equations give
rise to
$$\beta_1r_1f_{11}'+Y_1f_{11}+Z_1f_{12}=0\>,$$
$$\beta_1r_1f_{12}'-\beta_1r_1f_{12}+Y_2f_{12}+Z_2f_{11}=0\>,\eqno(28)$$
where prime signifies the differentiation with respect to $r_1$,
$Y_{1,2}=\alpha_1\beta_1\mp ga_1(\omega_1-gA_1)$,
$Z_{1,2}=\lambda_1\beta_1\pm ga_1\mu_0$. From (28) one yields the second order
equations in $r_1$
$$r_1f_{11}''+(1+2\alpha_1-r_1)f_{11}'-\frac{Y_1}{\beta_1}f_{11}=0\>,$$
$$r_1f_{12}''+(1+2\alpha_1-r_1)f_{12}'-
\left(1+\frac{Y_1}{\beta_1}\right)f_{12}=0\>,\eqno(29)$$
that are the Kummer equations (see, e. g. Ref.\cite{Abr64}) and their only
solution finite at 0 and at infinity not strongly increasing is the Laguerre
polynomial $L^\rho_{n_r^{(1)}}(r_1)$ (that always may be normalized
to be equal to 1 at $r_1=0$) with $\rho=2\alpha_1$
and $-n_r^{(1)}=Y_1/\beta_1=0,-1,-2,...$
If putting
$$f_{11}=C_{11}L^\rho_{n_r^{(1)}}(r_1),
f_{12}=C_{12}L^\rho_{n_r^{(1)}-1}(r_1)$$
then at $r_1=0$ from Eqs. (28) it follows
$C_{11}Y_1+C_{12}Z_1=0,\>C_{11}Z_2+C_{12}Y_2=0$ and since $Y_1Y_2-Z_1Z_2=0$
we obtain $C_{12}=-C_{11}\frac{Y_1}{Z_1}=-C_{11}\frac{Z_2}{Y_2}$. For
to describe relativistic bound states we should require
$\psi\in L_2^{12}({\Bbb R}^3)$ at any $t\in{\Bbb R}$ and one can
accept the normalization condition for
$F_{j1}, F_{j2}$ in the form
$$\int_0^\infty(|F_{j1}|^2+|F_{j2}|^2)dr=\frac{1}{3}\>\eqno(30)$$
with taking into account the condition (27) so that $C_{11}$ can be
determined from the relation (30). The condition
$Y_1/\beta_1=-n_r^{(1)}=0$ signifies $Y_1=0$ that entails $Z_1Z_2=0$. Then at
$\lambda_1>0$, $Z_1=2ga_1\mu_0>0$ (we consider $a_1>0$), $Z_2=0$, $C_{12}=0$ and
$n_r^{(1)}=0$ is admissible, i. e., $n_r^{(1)}=0,1,2...$.
At $\lambda_1<0$, $Y_1=Z_1=0$ but
$Y_2=2ga_1(\omega_1-gA_1)\ne0$, $Z_2=-2ga_1\mu_0\ne0$, $C_{12}\ne0$ and
$n_r^{(1)}=0$ is not allowed, i. e., $n_r^{(1)}=1,2,...$. Under the
situation the equality $Y_1/\beta_1=-n_r^{(1)}$ gives the spectrum
(provided that $\omega_1>gA_1>0$)
$$\omega_1=
gA_1+\mu_0\left[1+\frac{g^2a_1^2}{(n_r^{(1)}+
\sqrt{\lambda_1^2-g^2a_1^2})^2}\right]^{-1/2}\>.
\eqno(31)$$
  Further, if considering $a_2>0, \omega_3>gA_2>0$, $\omega_2<0$
with $|\omega_2|>g(A_1+A_2)$
then acting in line analogous with the above we have
$$\omega_3=gA_2+
\mu_0\left[1+\frac{g^2a_2^2}{(n_r^{(3)}+
\sqrt{\lambda_3^2-g^2a_2^2})^2}\right]^{-1/2},\>$$
$$\omega_2=-g(A_1+A_2)-\mu_0\left[1+\frac{g^2(a_1+a_2)^2}{(n_r^{(2)}+
\sqrt{\lambda_2^2-g^2(a_1+a_2)^2})^2}\right]^{-1/2},\>\eqno(32)$$
and the spectrum of meson is obtained in the form
$$\frac{\varepsilon}{\mu_0} =
\left[1+\frac{g^2a_1^2}{(n_r^{(1)}+
\sqrt{\lambda_1^2-g^2a_1^2})^2}\right]^{-1/2}
-\left[1+\frac{g^2(a_1+a_2)^2}{(n_r^{(2)}+
\sqrt{\lambda_2^2-g^2(a_1+a_2)^2})^2}\right]^{-1/2}$$
$$+\left[1+\frac{g^2a_2^2}{(n_r^{(3)}+
\sqrt{\lambda_3^2-g^2a_2^2})^2}\right]^{-1/2}\>,
\eqno(33)$$
where the numbers $n_{r}^{(2,3)}$ are subject to the same conditions as
for $n_r^{(1)}$.

\section{Influence of the Dirac-like monopole configurations of gluonic field}
If $K_j\ne0$ then according to the results of Refs.\cite{Gon99} we shall get
the similar relations providing that $K_j=k_j/g$ with $k_j\in{\Bbb Z}$,
i. e., $k_j$ are integer numbers. But now we should consider the spinor
$\Phi_j$ of (22) to be
the eigenspinor $\Phi_j$ of the twisted euclidean Dirac
operator ${\cal D}_k$ on the unit sphere ${\Bbb S}^2$, respectively, with
the Chern numbers
$k=k_1,-(k_1+k_2),k_2$ and the eigenvalues $\lambda_j$ should be, accordingly,
replaced by $\lambda_1=\pm\sqrt{(l_1+1)^2-k_1^2}$, $l_1\ge|k_1|$,
$\lambda_3=\pm\sqrt{(l_3+1)^2-k_2^2}$, $l_3\ge|k_2|$,
$\lambda_2=\pm\sqrt{(l_2+1)^2-(k_1+k_2)^2}$, $l_2\ge|k_1+k_2|$ (for more
details see Refs.\cite{Gon99}). Physically the corresponding configurations
of gluonic field describe the Dirac-like monopole ones with magnetic charges,
conformably, $P_1= k_1/g$, $P_2=-(k_1+k_2)/g$, $P_3=k_2/g$ but the total
(nonabelian) magnetic charge of the given configurations remains equal to
$P_1+P_2+P_3=0$. A number of authors (in particular, those who develop
lattice approach) insist on important role of such configurations in
a possible mechanism of the quark confinement (see, e. g. Refs.\cite{Mon}
and references therein).

\section{Spectrum of bound states in the Coulomb-linear case}
 We can get the approximate solutions of Eqs. (12)--(14) if considering
$\sigma_2\Phi_j=\Phi_j$ for the eigenspinor $\Phi_j$ of (22). As follows
from explicit form of those spinors found in Refs.\cite{Gon99} this can
be fulfilled only approximately and it is clear that when doing so we make
an error retaining eigenvalues $\lambda_j$ of the euclidean Dirac operator
${\cal D}_0$ on the unit sphere ${\Bbb S}^2$ instead of the eigenvalues
of a less
symmetric operator on ${\Bbb S}^2$ whose form is unknown explicitly. As will be
seen below, however, this may seemingly be compensated by the choice of
parameters $B_j$ of models so that the solutions obtained further may be
considered as the almost exact ones.
Let us employ the solutions of (16) and
(21) for the Yang-Mills equations (7) with the source $J$ described
late in Sec. 3 where at first we put $K_j=0$ as well. Using the given
solutions and the ansatz (22) we shall, in line with the above, get, e. g.
from (12), the system (23) which (having accepted
$\sigma_2\Phi_j\approx\Phi_j$) yields
$$\left[\partial_r+
\frac{\lambda_1}{r}-g\left(b_1+\frac{B_1}{r}\right)\right]F_{11}=
i(\mu_0-c_1)F_{12},$$
$$\left[\partial_r
-\frac{\lambda_1}{r}+g\left(b_1+\frac{B_1}{r}\right)\right]F_{12}=
-i(\mu_0+c_1)F_{11} \>.\eqno(34)$$
Now we employ the ansatz
$F_{11}=Ar^{\alpha_1}e^{-\beta_1r}[f_{11}(r_1)+f_{12}(r_1)]$,
$F_{12}=iBr^{\alpha_1}e^{-\beta_1r}[f_{11}(r_1)-f_{12}(r_1)]$
with $\alpha_1=\sqrt{(\lambda_1-gB_1)^2-g^2a_1^2}$, $\beta_1=
\sqrt{\mu_0^2-(\omega_1-gA_1)^2+g^2b_1^2}$, $A=gb_1+\beta_1$,
$B=\mu_0+\omega_1-gA_1$, $r_1=2\beta_1r$.
After this, inserting the ansatz into (34), adding and subtracting equations
entail
$$r_1ABf_{11}'+Y_1f_{11}+Z_1f_{12}=0\>,\eqno(35')$$
$$r_1ABf_{12}'-r_1ABf_{12}+Y_2f_{12}+\left(Z_2-\frac{gb_1}{\beta_1}
ABr_1\right)f_{11}=0\>,
\eqno(35'')$$
where prime signifies the differentiation with respect to $r_1$,
$Y_{1,2}=[\alpha_1\beta_1\mp ga_1(\omega_1-gA_1)+g\alpha_1b_1]B\pm g^2a_1b_1A$,
$Z_{1,2}=[(\lambda_1-gB_1)A\pm ga_1\mu_0)]B\pm g^2a_1b_1A$ and
$Y_1Y_2-Z_1Z_2=0$.

From (35$^{\prime}$--35$^{\prime\prime}$) one yields the second order
equations in $r_1$
$$r_1f_{11}''+(1+2\alpha_1-r_1)f_{11}'
+n_r^{(1)}f_{11}=0\>,\eqno(36)$$
$$r_1f_{12}''+\left(\frac{a}{a-br_1}+2\alpha_1-r_1\right)f_{12}'+
n_r^{(1)}\left(\frac{a\kappa}{a-br_1}+1\right)f_{12}=0\>\eqno(37)$$
with $a=Z_2$, $b=gb_1AB/\beta_1$, $\kappa=AB/Y_1$ and
$$n_r^{(1)}=\frac{gb_1Z_1-\beta_1Y_1}{\beta_1AB}.\eqno(38)$$
It is clear that according to (36) we should choose 
$f_{11}\sim L^\rho_{n_r^{(1)}}(r_1)$
with the Laguerre polynomial $L^\rho_{n_r^{(1)}}(r_1)$ ($\rho=2\alpha_1$)
if $n^{(1)}_r=0,1,2...$

A little bit below we shall show that one can choose
$f_{12}(r_1)\sim P^\rho_{n_r^{(1)}}(r_1)$ with some polynomial
$P^\rho_{n_r^{(1)}}(r_1)$, consequently,
we shall gain that
$\psi_1\in L_2^{4}({\Bbb R}^3)$ at any $t\in{\Bbb R}$ and, as a result,
we obtain from (38) that $\omega_1$ should be determined from the equation
$$[g^2a_1^2+(n_r^{(1)}+\alpha_1)^2](\omega_1-gA_1)^2+
2(\lambda_1-gB_1)g^2a_1b_1(\omega_1-gA_1)+$$
$$[(\lambda_1-gB_1)^2-(n^{(1)}_r+\alpha_1)^2]g^2b_1^2-
\mu_0^2(n^{(1)}_r+\alpha_1)^2=0\>,  \eqno(39)$$
that yields
$$\omega_1=gA_1-\frac{(\lambda_1-gB_1)g^2a_1b_1-
\sqrt{X_1}}{g^2a_1^2+(n^{(1)}_r+\alpha_1)^2}  \eqno(40) $$
with $X_1=(\lambda_1-gB_1)^2g^4a_1^2b_1^2-[g^2a_1^2+(n^{(1)}_r+\alpha_1)^2]
\{[(\lambda_1-gB_1)^2-(n^{(1)}_r+\alpha_1)^2]g^2b_1^2-
\mu_0^2(n^{(1)}_r+\alpha_1)^2\}$ and
the sign of $\sqrt{X_1}$ is chosen from the fact that the expression
(40) should pass on to (31) at $b_1=B_1=0$. The similar relations will also
hold true for $\omega_{2,3}$ with replacing
$a_1,A_1,b_1,B_1 \to a_2,A_2,b_2,B_2$ respectively for $\omega_3$ and
$a_1,A_1,b_1,B_1 \to -(a_1+a_2),-(A_1+A_2),-(b_1+b_2),-(B_1+B_2)$
respectively for $\omega_2$ and the spectrum of meson will be
$$\varepsilon=-\frac{(\lambda_1-gB_1)g^2a_1b_1-
\sqrt{X_1}}{g^2a_1^2+(n^{(1)}_r+\alpha_1)^2}$$
$$-\frac{[\lambda_2+g(B_1+B_2)]g^2(a_1+a_2)(b_1+b_2)+
\sqrt{X_3}}{g^2(a_1+a_2)^2+(n^{(2)}_r+\alpha_2)^2}
-\frac{(\lambda_3-gB_2)g^2a_2b_2-
\sqrt{X_2}}{g^2a_2^2+(n^{(3)}_r+\alpha_3)^2}\>,
\eqno(41) $$
where $\alpha_{2,3}$ and $X_{2,3}$ are obtained from $\alpha_1, X_1$ by
the same above replacements while
the signs of $\sqrt{X_j}$ are chosen to correspond to the expression
(33) in the Coulomb-like case $b_j=B_j=0$. We can see that all dependence of
$\lambda_j$ is expressed through the combinations $\lambda_j\pm gB_j$ and
the approximation error mentioned early in this section may probably be
compensated by choosing $B_j$ in a proper way.

Now let us describe the above polynomial $P^\rho_{n_r^{(1)}}(r_1)$
which should be the solution of the equation (37). Instead of solving (37),
however, we use the fact that the sought polynomial should also obey the
first order linear equation (35$^{\prime\prime}$) with $f_{11}(r_1)=
L^\rho_{n_r^{(1)}}(r_1)$ which allows us to write down
$P^\rho_{n_r^{(1)}}(r_1)$ in the form (putting $n_r^{(1)}\equiv n$,
$r_1 \equiv x$ for brevity)
$$P^\rho_n(x)=x^{-\alpha}e^{x}\int_0^{x}t^\alpha
\left(\frac{gb_1}{\beta_1}-\frac{Z_2}{ABt}\right)e^{-t}L^\rho_n(t)dt\>
\eqno(42)$$
with $\alpha=2\alpha_1+n-gb_1Z_1/(\beta_1AB)$. Further we can employ
the relation\cite{PBM2}
$$\int_0^x t^{\alpha-1}e^{-t}L^\rho_n(t)dt=
\frac{x^\alpha(1+\rho)_n}{n!\alpha}\,_2F_2(\alpha,1+\rho+n;
\alpha+1,1+\rho;-x)\>,{\rm Re}\,\alpha,x>0 $$
with the generalized hypergeometric function $_2F_2$ of order (2,2) while
the Pohgammer symbol $(a)_k=a(a+1)...(a+k-1)=\Gamma(a+k)/\Gamma(a)$
with the Euler gamma function $\Gamma(z)$, the relation\cite{PBM3}
$$ _2F_2(\alpha,1+\rho+n;
\alpha+1,1+\rho;-x)=\sum\limits_{k=0}^n\frac{(-1)^kx^k}{(1+\rho)_k}
C^k_n\frac{(\alpha)_k}{(\alpha+1)_k}\Phi(\alpha+k,\alpha+k+1;-x)$$
with the Kummer function $\Phi\equiv\,_1F_1$ and the relation\cite{Abr64}
$\Phi(z,z+1;-x)=zx^{-z}\gamma(z,x)$ with the incomplete gamma function
$\gamma(z,x)$ to come to the expression
$$ P^\rho_n(x)=\frac{(1+\rho)_n}{n!}x^{-\alpha}e^x
\left\{\sum\limits_{k=0}^n\frac{(-1)^k}{(1+\rho)_k}C^k_n
\left[\frac{gb_1}{\beta_1}\gamma(\alpha+k+1,x)-
\frac{Z_2}{AB}\gamma(\alpha+k,x)\right]\right\}\>. \eqno(43) $$

Now, using the functional equation $\gamma(z+1)=z\gamma(z)-x^ze^{-x}$,
it is not complicated to see that
$\gamma(\alpha+k+1,x)=(\alpha+k)\gamma(\alpha+k)-x^{\alpha+k}e^{-x}$,
$\gamma(\alpha+k,x)=(\alpha)_k\gamma(\alpha,x)-
[\sum_{i=1}^k(\alpha+i)_{k-i}x^{i-1}]x^\alpha e^{-x}$ and then one can show
with the help of (38) and of equality $Y_1Y_2-Z_1Z_2=0$ that
$$\frac{(1+\rho)_n}{n!}
\sum\limits_{k=0}^n\frac{(-1)^k}{(1+\rho)_k}C^k_n
\left[\frac{gb_1}{\beta_1}(\alpha+k)-
\frac{Z_2}{AB}\right](\alpha)_k=0\>. \eqno(44) $$
At last, employing the relation\cite{PBM1}
$$\frac{(1+\rho)_n}{n!}
\sum\limits_{k=0}^n\frac{(-1)^kx^k}{(1+\rho)_k}C^k_n=L^\rho_n(x)\>,$$
we come to the final form of the sought polynomial
$$P^\rho_0(x)=-\frac{gb_1}{\beta_1}L^\rho_0(x)=-\frac{gb_1}{\beta_1}\>,$$
$$P^\rho_n(x)=-\frac{(1+\rho)_n}{n!}
\sum\limits_{k=1}^n\frac{(-1)^k}{(1+\rho)_k}C^k_n
\left[\frac{gb_1}{\beta_1}(\alpha+k)-
\frac{Z_2}{AB}\right]\left[\sum\limits_{i=1}^k(\alpha+i)_{k-i}x^{i-1}\right]$$
$$-\frac{gb_1}{\beta_1}L^\rho_n(x)\>,n=1,2,3...\eqno(45) $$
Accordingly, we can now put $f_{11}=C_{11}L^\rho_n(x)$,
$f_{12}=C_{12}P^\rho_n(x)$ and from the system
(35$^{\prime}$--35$^{\prime\prime}$) at $r_1\equiv x=0$ it follows
the relation between the constants $C_{11},C_{12}$
$$C_{11}=-\frac{Z_1}{Y_1}\frac{P^\rho_n(0)}{L^\rho_n(0)}C_{12}=
-\frac{Y_2}{Z_2}\frac{P^\rho_n(0)}{L^\rho_n(0)}C_{12}\>.\eqno(46)$$
For completeness let us adduce the normalized radial part of 
$\psi_1$-component for
the meson wave function of ground state (where $n=0$ and
then $C_{11}=C_{12}=C)$
$$F_{11}=Cr^{\alpha_1}e^{-\beta_1r}\left(1-
\frac{gb_1}{\beta_1}\right)(gb_1+\beta_1)\>,$$
$$F_{12}=iCr^{\alpha_1}e^{-\beta_1r}\left(1+
\frac{gb_1}{\beta_1}\right)(\mu_0+\omega_1-gA_1)\>,\eqno(47)$$
where $C$ is determined from the relation (30) with the help
of formula\cite{PBM1}
$\int_0^\infty x^{\alpha-1}e^{-px}dx=\Gamma(\alpha)p^{-\alpha}$,
Re $\alpha,p >0$, which entails
$$C^2\left[\left(1-\frac{gb_1}{\beta_1}\right)^2(gb_1+\beta_1)^2+
\left(1+\frac{gb_1}{\beta_1}\right)^2(\mu_0+\omega_1-gA_1)^2\right]
\Gamma(2\alpha_1+1)(2\beta_1)^{-(2\alpha_1+1)}=\frac{1}{3}\>.\eqno(48)$$ 
It is evident that the similar expressions will hold true for the
corresponding radial parts of $\psi_{2,3}$-components with taking into account
the replacements described above.

Finally, it should be noted that the influence of the Dirac-like monopole 
configurations for gluonic field (when $K_j\ne0$) can be treated by the same 
manner as in Sec. 5 if taking $\sigma_2\Phi_j\approx\Phi_j$ for
the eigenspinor $\Phi_j$ of the twisted euclidean Dirac
operator ${\cal D}_k$ on the unit sphere ${\Bbb S}^2$ with the conforming
Chern numbers $k=k_1,-(k_1+k_2),k_2$.

\section{Concluding remarks}
  So we have seen that the black hole physics methods and results may really
help to build relativistic models of mesons
that are parametrized by constants $\mu_0, g, a_j, A_j, b_j, B_j, K_j$ which
should evidently be determined by comparing to experimental data. Even in the
case of pure Coulomb interaction these models give essentially different
spectrum of bound states in comparison with the naive picture of threefold
positronium-like spectrum accepted, e. g., in modern quarkonium spectroscopy.
It seems to us that the exact equations derived in the paper may serve as
a basis to construct miscellaneous approximate approaches in the spirit
of that of Sec. 6. We hope to continue further study of
the questions raised here.

\nonumsection{Acknowledgements}
\noindent
    The work was supported in part by the Russian Foundation for
Basic Research (grants Nos. 98-02-18380-a and 01-02-17157) and by
GRACENAS (grant No. 6-18-1997).

\nonumsection{References}

\end{document}